\documentclass{PoS}

\usepackage{amsmath}

\title{Origin(s) of Cosmic Rays}

\ShortTitle{Origin(s) of Cosmic Rays}

\author{\speaker{Luke O'C. Drury}\\
        Dublin Institute for Advanced Studies, 31 Fitzwilliam Place, Dublin 2, Ireland\\
        E-mail: \email{ld@cp.dias.ie}}

\abstract{The problem of the origin of Cosmic Rays is now over a century old and while there has been substantial progress, especially in the last decade, there are still open questions.  The question of "origin" is open to at least three possible interpretations depending on whether one follows the energy powering the accelerator, the matter being accelerated, or the physics of the acceleration process; these approaches are reviewed in turn. Supernova remnants remain by far the most plausible candidates as dominant sources for the bulk of the Galactic cosmic rays, but contributions from other source populations remain possible.  The transition at higher energies from Galactic to extra-galactic populations remains obscure.}

\FullConference{Cosmic Rays and the InterStellar Medium - CRISM 2014,\\
		24-27 June 2014\\
		Montpellier, France}

\begin{document}

\section{Introduction}

The discovery of cosmic rays is generally attributed to the Austrian scientist Viktor Hess \cite{Hess}, who was the first to clearly and unambiguously attribute the anomalous ionisation observed in the atmosphere to an "extremely penetrating radiation coming from above the atmosphere" on the basis of his balloon flights in 1912-1913.  Others had observed strange anomalies in ionisation, and in particular the Italian scientist Domenico Pacini \cite{Pacini} came very close to anticipating Hess on the basis of measurements at various depths under water.  Interestingly the suggestion was made as far back as 1900 by C T R Wilson \cite{Wilson} that the anomalous ionisation might be of extra-terrestrial origin, but this was not taken seriously until the work of Hess and its confirmation by Kolh\"orster \cite{Kolhoerster} in 1914.

Remarkably, for a phenomenon that was discovered over a century ago, the origin of Cosmic Rays remains an active field of research.
It is worth noting that the phrase "origin of cosmic rays" is open to at least three somewhat different interpretations (see \cite{Drury12} for a similar discussion on which this presentation is largely based).  We can follow the energy and ask where the power comes from to drive the acceleration.  We can follow the matter and ask what is the source of the material being accelerated.  And perhaps most physically relevant, we can ask how and where the acceleration occurs.  These three approaches will hopefully converge to a consistent picture.  

\section{Following the energy}

How much power is required to maintain the observed Galactic cosmic ray flux?  The conventional estimate, derived from rather robust arguments relating to the amount of matter traversed by the cosmic rays in the interstellar medium as deduced from the production of secondary spallation nuclei, is $10^{41}\,\rm erg\, s^{-1}$ or $10^{34}\,\rm W$.  The classic monograph by Ginzburg and Syrovatskii \cite{GiSy} gives a conservative estimate of $0.3\times10^{34}\,\rm W$, while in a recent paper the authors of Galprop quote a remarkably precise value of $(0.7\pm0.1)\times 10^{34}\,\rm W$.  At the higher end Drury, Markiewicz and V\"olk \cite{DMV} suggest $3\times 10^{34}\,\rm W$.  The differences between these estimates relate largely to the choice of propagation model assumed and are thus in a sense systematic and not statistical errors.  There are two issues.  Firstly, how hard is the true injection spectrum?  The high estimate of Drury, Markeiwicz and V\"olk results from assuming an injection spectrum with an $E^{-2}$ energy dependence whereas the lower estimates assume softer injection spectra more like $E^{-2.3}$ ($E$ is of course particle energy).  The second issue is how much energy is contributed at low energies by second order Fermi acceleration in propagation models assuming re-acceleration.  This last is an interesting issue which has received surprisingly little attention, but which may well be very significant.  In \cite{TD} we show that the re-acceleration power, which ultimately derives from the damping of interstellar turbulence on low-energy cosmic rays, can be written as

\begin{eqnarray}
P_R &=& \int_0^\infty 4\pi p^2 f {1\over p^2} {\partial\over\partial p} \left(p^4V_A^2 v\over 9 D_{xx}\right) \, dp\\
&=& \int_0^\infty 
4\pi p^2 f \left (V_A^2 p v\over 9 D_{xx}\right) \left[ 4 + {\partial\ln(v/D_{xx})\over\partial\ln p}\right].
\end{eqnarray}
In this expression  $p$ denotes particle momentum, $v$ particle velocity, $f$ the isotropic part of the particle's phase space density, 
$V_A$ is the Alfv\'en speed and $D_{xx}$ is the spatial diffusion coefficient of the cosmic rays.  As assumed in, for example, the Galprop propagation code, the second-order acceleration is related to the spatial diffusion coefficient by supposing that both result from scattering of particles by Alfv\'enic turbulence.
We note that $pv = T$ for relativistic particles and $2T$ for non-relativistic particles where $T$ is the kinetic energy and the logarithmic slope term is small for any reasonably power-law dependence of $D_{xx}$ on $p$,
\begin{equation}
\left|{\partial\ln(v/D_{xx})\over\partial\ln p}\right| <1.
\end{equation}
In fact with the conventional parametrisation $D_{xx}\propto v p^\delta$ the final bracket is just $[4-\delta]$ and in the generally favoured reacceleration models $\delta = 0.3$.

Thus to order of magnitude the re-acceleration time scale, 
\begin{equation}
{E_{CR}\over P_R} \approx {9\over 4} {D_{xx}\over V_A^2}
\end{equation}
as was to be expected on dimensional grounds.

Another interesting way to write the above expression is in terms of the scattering time $\tau$.  Substituting $D_{xx}=v^2\tau/3$ we get
\begin{equation}
P_R = \int_0^\infty 
4\pi p^2 f \left (V_A^2 p v\over 3 v^2 \tau\right) \left[ 4 + {\partial\ln(v/D_{xx})\over\partial\ln p}\right]
\end{equation}
which can be interpreted as saying that the average energy gain per scattering of each particle is just
\begin{equation}
\Delta T = pv {4\over 3} {V_A^2\over v^2} \left[ 1 + {1\over 4}{\partial\ln(v/D_{xx})\over\partial\ln p}\right].
\label{deltaT}
\end{equation}
Clearly significant amounts of energy can be transferred if there are large numbers of sub-relativistic cosmic rays and/or strong interstellar turbulence.  This is not normally taken into account.

In summary, taking into account the systematic uncertainties due to the propagation model dependencies it seems certain that the Galactic cosmic ray luminosity must lie in the range 
\begin{equation}
0.3\times 10^{34}{\,\rm W} < L_{GCR} < 3\times 10^{34}{\,\rm W}.
\end{equation}
and probably closer to the upper bound than the lower.
As is well know the mechanical energy input into the interstellar medium by Supernova explosion is of order $10^{35}\,\rm W$ so that supernovae are a possible power source for the acceleration if the acceleration process can operate at relatively high efficiency.  In fact there is no other plausible known source of enough energy in the Galaxy although pulsars and OB winds may contribute at the 10\% level.

\begin{figure}[htbp]
\begin{center}
\includegraphics[width=\textwidth]{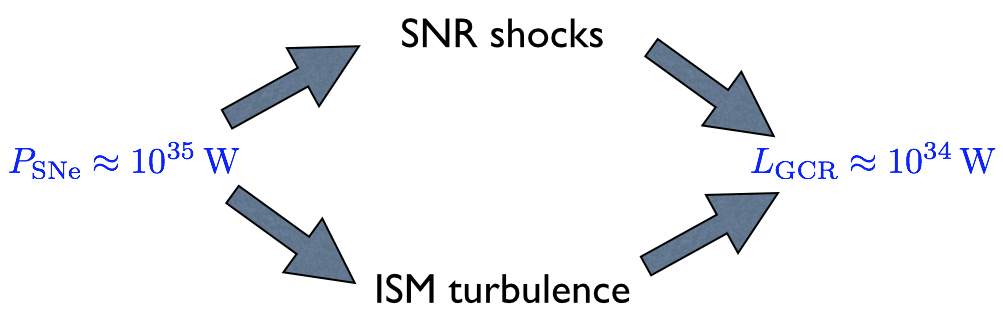}
\caption{Following the energy.}
\label{F1}
\end{center}
\end{figure}

This has long been seen as a strong hint that the ultimate source of the energy driving the acceleration of the Galactic cosmic rays must be sought in supernovae.  However the acceleration cannot take place directly in the explosion itself because there would be strong adiabatic losses in the expansion of the subsequent Supernova remnant (SNR).  Rather the acceleration must be mediated through shocks and/or turbulence driven by the SNRs as they expand into the interstellar medium (ISM) as schematically shown in Fig~\ref{F1}.

It is amusing to note that the Solar wind is a cosmic ray accelerator at low energies (Solar modulation is just the expulsion of low energy cosmic rays from the solar system by the Solar wind, which must therefore be doing work on the cosmic rays) but the power involved is negligible \cite{Drury12}, even if integrated over all the solar-type stars in the Galaxy.

\section{Following the matter}

An alternative approach is to ask where the matter comes from that is fed into the accelerator.  This can be done by looking for chemical and isotopic signatures and interpreting them in terms of possible sources.  At low to moderate energies such studies are technically quite easy and there is now an extensive body of data on the chemical and isotopic composition of the GCR at energies around a few GeV per nucleon.  It is of course necessary to correct the observed abundances for the effects of nuclear interactions during propagation, but this is relatively straightforward.

The bottom line is that all the chemical elements up to and including Uranium are seen in the GCR, and to a first approximation they all have essentially identical power-law spectra.  Recently it has become clear that this simple picture needs some second-order corrections and that Helium in particular appears to have a slightly harder spectrum than the protons. This does not seriously affect the discussion of chemical composition by which we mean the relative numbers of different nuclei observed in the GCR at fixed energy per nucleon.   After correction for spallation during propagation the general pattern of the chemical composition is quite normal.  The major nucleosynthetic features are evident, a peak around CNO and another at Fe and a rapid decrease for charges higher than Fe.  To proceed further it is necessary to have some reference pattern of abundances against which to compare the observations and the common practice is to use standard Solar system abundances as a reference.  These seem to be a good proxy for the bulk composition of the ISM and are well determined from meteoritic and other studies. Relative to this "standard" abundance pattern a striking feature of the GCR is that many of the heavier elements, such as Fe, SI and Ca, are over-abundant by factors of about 30 and that the data do not seem to be easily organised in any one-parameter model.  There is certainly a general tendency for heavier species to be over-abundant, but factors other than atomic weight are clearly involved and there are strong hints that this second factor is related to chemistry.

Historically the first attempt to explain the data invoked an {\it ad hoc} dependence on the first ionisation potential (FIP) of each element in a so-called FIP-bias.  Of course the first ionisation potential is intimately related to the chemical properties of each element (which are determined by the outer electronic structure of the neutral atom) and one can equally well explain the data with a volatility dependent fractionation.  The most detailed and physically motivated attempt at interpreting the compositional data is based on the idea that in most of the ISM the refractory species are not in the gas phase but in dust grains.  In \cite{MDE, EDM} the data are interpreted in terms of standard shock acceleration operating in a dusty ISM with charged dust grains which are slightly accelerated and then partially sputtered upstream of the shock.  The basic idea is that there are two routes for ions to enter the accelerator.  Gas phase species, such as the noble gases, have to be injected by plasma physics processes operating in the shock which exhibit a strong mass to charge fractionation favouring heavier species; elements mainly in the solid dust phase are sputtered off accelerated dust grains in the upstream region as already supra-thermal ions and swept into the shock where they are accelerated with relatively little fractionation.  This is shown in \cite{EDM} to offer a good explanation of the data.  In particular, the fact that the oxygen over-abundance lies between the pure gas-phase and pure refractory elements is well explained by the fact that about 10\% of the interstellar oxygen should be locked up in grains (the heavy metals typically form silicate minerals incorporating a lot of oxygen).   

The overabundances of the heavy elements can be somewhat reduced by supposing that the starting material has higher metallicity than Solar, as should be the case inside a super bubble for example, but the general pattern remains the same and the arguments for a signature of dust chemistry remain \cite{Rauch09}.  The consensus thus is that the overall elemental abundances point to an origin in a rather normal, well-mixed and dusty ISM.  It is worth pointing out that the composition requires the normal mixture of nucleosynthetic components and is not compatible with the ejecta of any one class of SNe.

There is evidence of a recent admixture of some relatively freshly synthesised material in the ultra-heavy element abundances.  The largest sample is that reported from the Ultra-Heavy Cosmic Ray Experiment (UHCRE) which flew on NASA's Long Duration Exposure Facility \cite{Donnelly} and accumulated an exposure of $170\,\rm m^2sr\,yr$ but with poor charge resolution.  This saw some 35 Actinide events including one possible trans-Uranic Curium nucleus.   Thorium appears to be significantly more abundant than Uranium, pointing to a relatively old population where significant decay of Uranium relative to Thorium has occurred; this is of course in contradiction to the observation of a short-lived Curium event (if this is real) and suggests that what we are seeing is mainly old material but with some fresh contamination at the level of a few percent.   The one well-established isotopic anomaly in the GCRs is a clear overabundance of $^{22}$Ne which again is consistent with some admixture of Wolf-Rayet star wind material as might be expected in a super-bubble or large OB association.

\begin{figure}[htbp]
\begin{center}
\includegraphics[width=\textwidth]{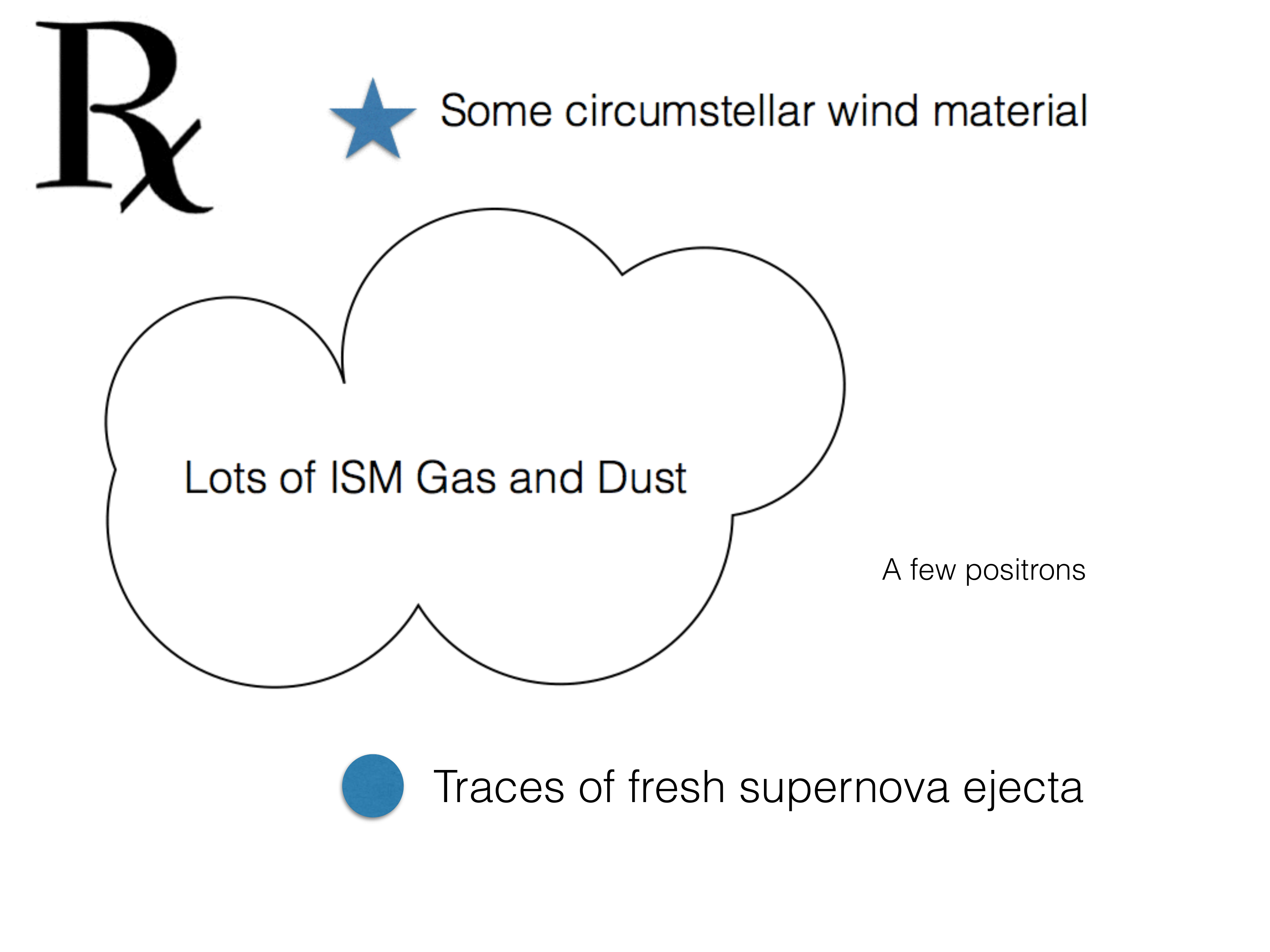}
\caption{Following the matter; a prescription for GCR composition.}
\label{F2}
\end{center}
\end{figure}

In conclusion the compositional evidence points to an origin of the nuclear matter in rather normal dusty ISM material with some recent contamination from nucleosynthesis in massive stars and SNe. This is shown graphically in Fig.~\ref{F2}.  It should be noted that the recent detection by Pamela \cite{PAMELA} and AMS of a clear positron excess at high energies points to at least one additional source of high-energy positrons (and presumably electrons), plausibly a contribution from near-by pulsar wind nebulae.

\section{Following the Physics}

On the theoretical front the leading candidate (indeed almost the only candidate) mechanism for the acceleration is the so-called Diffusive Shock Acceleration (DSA) first described by \cite{Krymsky} in a peer-reviewed publication.  The same process was almost simultaneously described by \cite{Axford, Bell, BO} and is in essence a fast and efficient variant of Fermi acceleration in which charged particles are accelerated by the compression of the flow in the shock front.  As an acceleration mechanism for cosmic rays DSA has a number of advantages.  Firstly, there is no need for a separate "injection" process; DSA can operate right down to mildly supra-thernal energies so that the GCRs emerge naturally as an extended non-thermal tail of the shock-heated ion distributions.  Secondly, it naturally and without artificial fine-tuning produces approximate power-law spectra with exponents close to $E^{-2}$. Thirdly it can be efficient and convert a significant part of the energy dissipated in the shock into non-thermal particle energy.  Recent progress in computer simulations 
\cite{CS1, CS2, CS3, CPS} has largely confirmed our theoretical expectations developed over the last few decades and demonstrated efficient injection and acceleration of ions at non-relativistic shocks accompanied by strong magnetic field amplification and wave excitation leading to Bohm diffusion (mean free path of order the gyro-radius).  A word of caution is however in order.  Despite the very impressive advances in the simulations, these are still restricted in their dynamic range and do not as yet extend to relativistic energies and strongly non-linear shock modification.  

The amplification of the magnetic field, as suggested by Bell, \cite{Bell, DD}, together with the fact that the particle diffusion is in the Bohm limit, is of great importance and represents a significant advance in our theoretical understanding.  It is easy to show that if the diffusion is driven to the Bohm limit, then the maximum particle rigidity to which a particle can be accelerated at a shock of radius $R$ expanding at velocity $\dot R$ is of order
\begin{equation}
B R \dot R
\end{equation} where $B$ is the magnetic field strength.
This is the well know Lagage and Cesarsky \cite{LC} limit and is a special case of the general Hillas limit \cite{Hillas}.  It can be easily seen to follow from simple dimensional arguments.  Multiplying a magnetic field by a velocity gives an electric field, and multiplying this by a length scale we get a potential (in volts if we use SI units).  More detailed estimates usually give a numerical factor of order $10^{-1}$ in front of this expression, and inserting typical values for a supernova remnant and a standard interstellar magnetic field strength Lagage and Cesarsky concluded that the maximum particle rigidity fell some way short of the ``knee'' feature in the cosmic ray spectrum.  

For a supernova remnant the product $R\dot R$ is expected to rise to a maximum at the start of the Sedov phase and then slowly decrease.  In the Sedov phase the radius grows with time as $R\propto t^{2/5}$ and in consequence the shock velocity drops as $\dot R \propto t^{-3/5}$, thus $R\dot R \propto t^{-1/5}$.  The radius at the start of the Sedov phase is of order the mass sweep-up radius where the amount of ambient matter (of density $\rho_0$) swept up by the shock is equal to that ejected in the explosion $M_{\rm ej}$.
\begin{equation}
{4\over 3} \pi R^3 \rho_0 = M_{\rm ej} \implies R\propto \left(M_{\rm ej}\over\rho_0\right)^{1/3}
\end{equation}
Similarly the shock velocity at the start of the Sedov phase is determined by the ratio of the explosion energy $E_{\rm SN}$ to the ejecta mass $E_{\rm ej}$,
\begin{equation}
E_{\rm SN} \approx {1\over 2} M_{\rm ej} \dot R^2 \implies \dot R \propto \left (E_{\rm SN} \over M_{\rm ej}\right)^{1/2}
\end{equation}
and thus 
\begin{equation}
R\dot R \propto E_{\rm SN}^{1/2} M_{\rm ej}^{-1/6} \rho_0^{-1/3} t^{-1/5}
\end{equation}
with rather weak dependence on all parameters.  The only hope therefore if we want to substantially increase $B R \dot R$ is to amplify the magnetic field $B$.  It should be noted that the field has to be amplified on {\em both} sides of the shock, and on scales that can interact with the highest energy particles; this appears to be possible.

In addition to the theoretical attraction of magnetic field amplification for accelerating high energy ions, as required if one want to make the GCRs at least as far as the ``knee'', there is now strong observational evidence coming from X-ray observations of synchrotron emission from very high energy electrons.  This produces extremely thin hard X-ray rims at the outer shock of young remnants.  A recent excellent observational study is that of Ressler {\it et al} \cite{Ressler} who study the rims of SN1006 in some detail.  They conclude that the magnetic field is amplified by factors of at least 10 to 50 and that there is evidence for very small diffusion coefficients, even down to the Bohm limit.

With the addition of field amplification (the last missing piece of the jig-saw) DSA at SNRs appears to be capable of producing the bulk of the GCRs at least as far as the ``knee'' energy of $3\times 10^{15}\,\rm eV$ with a tail extending to higher energies from some exceptional SNRs in an early phase.   Making the effective magnetic field strength depend on the shock speed has the effect of moving high-energy production to earlier times, and of making particle escape a much more important aspect of the process \cite{Escape}.  Exactly how and where the transition from a Galactic dominated population to an extra-galactic component occurs is however obscure.  Observationally while the all-particle energy spectrum is now very well determined, attempts to separate this into spectra for different nuclear species are extremely challenging and the picture is quite confused.  Hopefully this will improve over the next few years with better nuclear interaction models and more data.

\begin{figure}[htbp]
\begin{center}
\includegraphics[width=\textwidth]{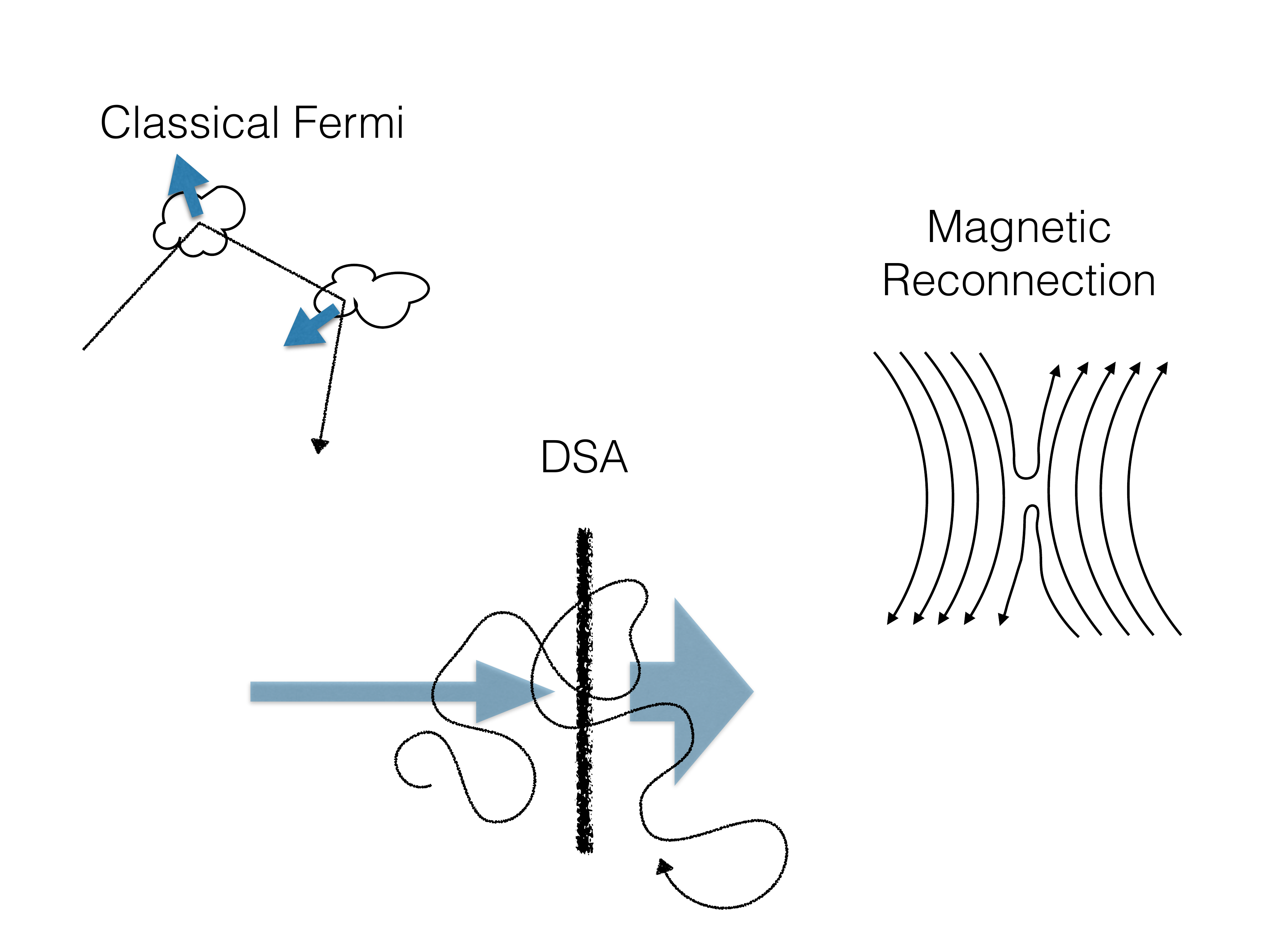}
\caption{Following the Physics.  Cartoon representations of three possible acceleration mechanisms}
\label{F3}
\end{center}
\end{figure}

While the bulk of the strong shocks in the Galaxy are those driven by supernovae, there are other possibilities which should not be forgotten, such as interacting stellar winds in OB associations or super-bubbles.  These could and indeed should contribute to the production of the GCR; DSA should operate at any sufficiently strong shock running into a tenuous plasma.  Also DSA is not the only possibility.  At low energies second order Fermi on interstellar turbulence may be a significant contributor (as argued above) and magnetic reconnection is a viable alternative to DSA if enough energy can be put into winding up the magnetic field.

\section{Conclusions}

The evidence from following the energy, the composition and the physics is consistent with a picture where the bulk of the GCR are accelerated by the DSA mechanism at the strong shocks driven by SNe into the ISM.  There is however room for other contributions at the 10\% level and indeed the PAMELA positron excess clearly shows that at least for the high energy electrons an additional source is required.


\begin{thebibliography}{99}

\bibitem{Hess} V. F. Hess, \it \"Uber Beobachtungen der durchdringenden Strahlung bei sieben Freiballonfahrten, \it Phys. Z. \bf 13 \rm 1084 (1912).

\bibitem{Pacini} D. Pacini, \it La radiazione penetrante alla superficie ed in seno alle acque, Nuovo Cim. \bf VI/3 \rm 93 (1912)

\bibitem{Wilson} C. T. R. Wilson, \it On the Leakage of Electricity through Dust-free Air, Proc. Camb. Soc. \bf 11 \rm 32 (1900)

\bibitem{Kolhoerster} W. Kolh\"orster \it Messungen der durchdringenden Strahlung im Freiballon in gr\"o{\ss}eren H\"ohen, Phys. Z. \bf 14 \rm 1153; \it Ber. deutsch. Phys. Ges. \bf 16 \rm 719 (1914)

\bibitem{Drury12} L. O'C. Drury, \it Origin of Cosmic Rays, Astroparticle Phys. \bf 39 \rm 52 (2012) [arXiv:1203.3681]

\bibitem{GS} V. L. Ginzburg and S. I. Syrovatskii, \it The Origin of Cosmic Rays, \rm authorised English translation by H. S. H. Massey, Pergamon Press, Oxford (1964).

\bibitem{DMV} L. O'C. Drury, W. Markiewicz and H. J. V\"olk, \it Simplified models for the evolution of supernova remnants including particle acceleration, A\&A \bf 255 \rm 179 (1989)

\bibitem{TD} A. Thornbury and L. O'C. Drury, \it Power requirements for cosmic ray propagation models involving re-acceleration and a comment on second-order Fermi acceleration theory, MNRAS \bf 442 \rm 3010 (2014) [arXiv:1404.2104]

\bibitem{MDE}{J-P. Meyer, L. O'C. Drury and D. C. Ellison, \it Galactic Cosmic Rays from Supernova Remnants. I. A Cosmic-Ray Composition Controlled by Volatility and Mass-to-Charge Ratio, ApJ \bf 487 \rm 182 (1997} [arXiv:9704267]

\bibitem{EDM} D. C. Ellison, L. O'C. Drury and J-P. Meyer, \it Galactic Cosmic Rays from Supernova Remnants. II. Shock Acceleration of Gas and Dust, ApJ \bf 487 \rm 197 (1997) [arXiv:970493]

\bibitem{Rauch09} B. F. Rauch {\it et al}, \it Cosmic Ray origin in OB Associations and Preferential Acceleration of Refractory Elements: Evidence from Abundances of Elements 26Fe through 34Se, ApJ \bf 697 \rm 2083 )2009) [arXiv:0906.2021]

\bibitem{Donnelly}J. Donnelly \it et al, Actinide and Ultra-Heavy Abundances in the Local Galactic Cosmic Rays: An Analysis of the Results from the LDEF Ultra-Heavy Cosmic-Ray Experiment, ApJ \bf 747 \rm 40 (2012)

\bibitem{PAMELA} O. Adriani \it et al, Cosmic-Ray Positron Energy Spectrum Measured by PAMELA, 
PRL, \bf 111, \rm 081102 (2013) [arXiv:1308.0133]

\bibitem{Krymsky} G. F. Krymsky, \it A regular mechanism for the acceleration of charged particles on the front of a shock wave, 
Akademiia Nauk SSSR Doklady, \bf 234, \rm 1306 (1977)

\bibitem{Axford} W. I. Axford, E. Leer and G. Skadron, \it The acceleration of cosmic rays by shock waves, Proc. 15th ICRC (Plovdiv) \bf 11 \rm 132 (1978)

\bibitem{Bell} A. R. Bell, \it The acceleration of cosmic rays in shock fronts. I, MNRAS \bf 182 \rm 147 (1978)

\bibitem{BO} R. D. Blandford and J. P. Ostriker, \it Particle acceleration by astrophysical shocks, ApJ \bf 221 \rm L29 (1978)

\bibitem{CS1} D. Caprioli and A. Spitkovsky, \it Simulations of Ion Acceleration at Non-relativistic Shocks. I. Acceleration Efficiency,
ApJ \bf 783 \rm 91 (2014) [arXiv:1310.2943]

\bibitem{CS2} D. Caprioli and A. Spitkovsky, \it Simulations of Ion Acceleration at Non-relativistic Shocks. II. Magnetic Field Amplification,
ApJ \bf 794 \rm 46 (2014) [arXiv:1401.7679]

\bibitem{CS3} D. Caprioli and A. Spitkovsky, \it Simulations of Ion Acceleration at Non-relativistic Shocks. III. Particle Diffusion,
ApJ \bf 794 \rm 47 (2014) [arXiv:1407.2261]

\bibitem{CPS} D. Caprioli, A. Pop and A. Spitkovsky, \it  Ion Injection at Non-relativistic Collisionless Shocks,\rm [arXiv:1409.8291]

\bibitem{LC} P. O. Lagage and C. Cesarsky, \it The maximum energy of cosmic rays accelerated by supernova shocks, 
A\&A, \bf 125, \rm 249 (1983)

\bibitem{Hillas} A. M. Hillas, \it The Origin of Ultra-High-Energy Cosmic Rays, Annual Review of Astronomy and Astrophysics, \bf 22, \rm
425 (1984)

\bibitem{Bell04} A. R. Bell, \it Turbulent amplification of magnetic field and diffusive shock acceleration of cosmic rays
MNRAS \bf 353 \rm 550 (2004) 

\bibitem{DD} T. P. Downes and L. O'C. Drury, \it Cosmic ray pressure driven magnetic field amplification: dimensional, radiative and field orientation effects, MNRAS \bf 444 \rm 365 (2014) [arXiv:1407.5664]

\bibitem{Ressler} S. M Ressler \it et al, Magnetic Field Amplification in the Thin X-Ray Rims of SN 1006, ApJ \bf 790 \rm 85 (2014)
[arXiv:1406.3630]

\bibitem{Escape} L. O'C. Drury, \it Escaping the accelerator: how, when and in what numbers do cosmic rays get out of supernova remnants?
MNRAS \bf 415 \rm 1807 (2011) [arXiv:1009.4799]

\end{thebibliography}
\end{document}